\documentclass[preprint, letterpaper, nobibnotes, aps, superscriptaddress,prb]{revtex4-1}

\usepackage{pdfpages}
\usepackage{etoolbox}
\usepackage{graphicx}
\usepackage{amssymb}
\usepackage{amsmath}
\usepackage{color}
\usepackage{url}
\usepackage{multirow}
\usepackage{graphicx}        
\graphicspath{ {../figures/} }
\usepackage{bm}
\usepackage[justification=centering]{caption}
\usepackage{float}
\usepackage{boldline}
\usepackage{mciteplus}
\usepackage[version=3]{mhchem} 
\usepackage[T1]{fontenc}       

\newtoggle{fulltitlepage}
\settoggle{fulltitlepage}{true}

\begin{document}

\title{Predicting Solvation Free Energies and Thermodynamics in Polar Solvents and Mixtures Using a Solvation-Layer Interface Condition}

\iftoggle{fulltitlepage}{
\author{Amirhossein Molavi Tabrizi}
\affiliation{Department of Mechanical and Industrial Engineering, Northeastern University, Boston, MA, USA}

\author{Spencer Goossens}
\affiliation{Department of Mechanical and Industrial Engineering, Northeastern University, Boston, MA, USA}

\author{Ali Mehdizadeh Rahimi}
\affiliation{Department of Mechanical and Industrial Engineering, Northeastern University, Boston, MA, USA}

\author{Matthew Knepley}
\affiliation{Department of Computational and Applied Mathematics, Rice University, Houston, TX, USA}

\author{Jaydeep P. Bardhan}
\affiliation{Department of Mechanical and Industrial Engineering, Northeastern University, Boston, MA, USA}
\email{j.bardhan@neu.edu}

\begin{abstract}
We demonstrate that with two small modifications, the popular dielectric continuum model is capable of predicting, with high accuracy, ion solvation thermodynamics in numerous polar solvents, and ion solvation free energies in water--co-solvent mixtures.  The first modification involves perturbing the macroscopic dielectric-flux interface condition at the solute--solvent interface with a nonlinear function of the local electric field, giving what we have called a solvation-layer interface condition (SLIC).  The second modification is a simple treatment of the microscopic interface potential (static potential). We show that the resulting model exhibits high accuracy without the need for fitting solute atom radii in a state-dependent fashion.  Compared to experimental results in nine water--co-solvent mixtures, SLIC predicts transfer free energies to within 2.5~kJ/mol.  The co-solvents include both protic and aprotic species, as well as biologically relevant denaturants such as urea and dimethylformamide.   Furthermore, our results indicate that the interface potential is essential to reproduce entropies and heat capacities.  The present work, together with previous studies of SLIC illustrating its accuracy for biomolecules in water, indicates it as a promising dielectric continuum model for accurate predictions of molecular solvation in a wide range of conditions.
\end{abstract}}
{}


\maketitle

\section{Introduction}




Developing better models for thermodynamics of solute--solvent interactions is of crucial importance due to their wide range of applications in biology, nanotechnology, and fundamental chemistry. Implicit-solvent models play a variety of roles in these applications because their speed and simplicity make them appealing options in applications where fully atomistic explicit-solvent models are impractical or impossible~\cite{Sharp90,Roux99}. Among the most popular implicit-solvent models are those based on  statistical mechanical integral equations~\cite{Kovalenko98,Luchko10,Remsing16} and those based on macroscopic dielectric theory and continuum electrostatics~\cite{Kirkwood34,Sharp90,Tomasi94}. The latter are widely used because they lead to well understood partial-differential equations for which a variety of numerical algorithms can be used to solve large problems~\cite{Gilson85,Honig86,Gilson88,Honig95,Fan05,Carlsson06,Bertonati07}.

However, the speed advantage of dielectric models comes at the cost of simplifying assumptions that make them unable to capture important phenomena~\cite{Latimer39,Rashin85_Honig,Ashbaugh00,LyndenBell01,Rajamani04,Fedorov07,Mobley08_asymmetry,Fennell11,Bardhan12_asymmetry,Shi13,Bardhan14_asym,Bardhan15_PIERS}. In particular, the most substantial errors are incurred in the continuum theory's treatment of the first layers of solvent molecules (the solvation layer) as bulk dielectric material.  Significant inaccuracies arise from the assumptions that solvent molecules (1) are infinitely small, and (2) respond linearly with respect to an applied field~\cite{Beglov96,Roux99}.  To understand the behavior of solvent molecules in this layer, numerous groups have assessed physically motivated changes to solute atom radii~\cite{Latimer39,Rashin85_Honig} and conducted all-atom calculations with explicit solvent to probe solvation-layer response to a perturbing electric field~\cite{Ashbaugh00,Rajamani04,Grossfield05,Hummer96,Jayaram89,Alper90,Remsing16}.

These studies, which integrate extensive experimental and computational data, have supported the development of several dielectric-based models that address solvation-layer phenomena for water~\cite{Rashin85_Honig,Purisima09,Corbeil10,Fawcett93,Chamberlin08,Mukhopadhyay14,Duignan14,Reif16}. Many focus on charge hydration asymmetry (CHA), that is, reproducing the fact that ions of equal size but opposite valence have different solvation free energies and entropies.  Although existing models have provided improved treatment of CHA, they have generally treated all asymmetry as arising solely from water hydrogens approaching a solute more closely than the larger water oxygens. This phenomenon is known as steric asymmetry. In many continuum models, steric asymmetry is addressed using atom-type-specific or charge-dependent radii~\cite{Latimer39,Rashin85_Honig,Nina97,Fawcett92,Fawcett04,Purisima09,Mukhopadhyay14,Sundararaman15}. Although effective radii do account for the effects of charge asymmetry, the fact that the electric field will be disturbed by a buried charge suggest that this correction should be applied to the interface rather than to the atom radii directly\cite{Bardhan12_asymmetry,Bardhan14_asym,Bardhan15_PIERS,Molavi16}.  Compounding the challenge of modeling asymmetric response is that it is better described as a combination of two distinct different mechanisms\cite{Bardhan12_asymmetry}, one being the steric asymmetry, and the other being
an electrostatic interface potential that persists even if the solute is uncharged~\cite{Ashbaugh00,Rajamani04,LyndenBell01,Cerutti07,Kathmann11,Lin14}.  This interface potential, which we call a static potential to distinguish it from the macroscopic notion~\cite{Bardhan12_asymmetry}, contributes substantially to solvation thermodynamics, though not to solvation free energies in the case of neutral solutes.  In particular, the static-potential term contributes a term that is linear in the net charge~\cite{Bardhan12_asymmetry,Reif16}, whereas the polarization contributes the familiar quadratic expression. For linear-response models this quadratic dependence is clearly understood, and for our nonlinear response model it arises from the fact that the model responds linearly for virtually the entire charging process.

We have proposed a corrected dielectric continuum model that includes two simple modifications to treat these phenomena directly and separately~\cite{Bardhan12_asymmetry,Bardhan14_asym,Bardhan15_PIERS,Molavi16_JCTC}.  First, the static potential is treated as a uniform field that does not change in response to the solute charge distribution; second, we modify the familiar dielectric flux interface condition (obtained from macroscopic dielectric theory) by adding a nonlinear perturbation that depends on the local electric field. We call this the solvation-layer interface condition (SLIC) model, after the modified interface condition~\cite{Molavi16}.  Our initial work showed that SLIC accurately reproduces ion solvation free energies in water, as well as charge-hydration asymmetries on a challenging test set~\cite{Bardhan14_asym}.  We then established that the widely used mean spherical approximation (MSA) in bulk solution theory~\cite{Blum87} could be approximated to give a SLIC-like nonlinear perturbation to the macroscopic dielectric interface condition~\cite{Molavi16}; this work indicated that a temperature-dependent interface condition could accurately predict solvation free energies and entropies in a variety of polar solvents~\cite{Molavi16}.  Most recently, SLIC has been extended for dilute electrolytes modeled with the linear Poisson-Boltzmann equation~\cite{Bardhan15_PIERS,Molavi16_JCTC}. This extended version was shown to accurately predict the charging free energies of individual atoms in polyatomic solutes~\cite{Molavi16_JCTC}.  Remarkably, the model provides high accuracy without the need for parameterizing solute atom radii.

In this paper, we test the SLIC model on two problems that are widely understood to challenge traditional dielectric continuum models. First, it is well known that such models fail to reproduce solvation thermodynamics~\cite{Vath99}; the problem's importance has in fact motivated the parameterization of temperature-dependent radii~\cite{Elcock97}.  Second, relatively few implicit-solvent models have been applied to solvation in mixtures~\cite{Allen09,Basilevsky09,Bonincontro06,Nakamura12}. Standard dielectric models have been shown to give poor accuracy in specific mixtures~\cite{Jarzeba91,Kolling91,Barnes15}, but reference-interaction site model (RISM) theories~\cite{Yoshida02} and the continuum-based model COSMO-RS~\cite{Klamt01,Klamt11,Klamt15} generally work well.  One challenge for simple dielectric models is that correcting their oversimplifications, even in pure solvents, necessitates numerous correction terms with associated free parameters, making parameterization prohibitively complicated and time-consuming.  For the studies here, where we use standard Shannon-Prewitt radii for the ions~\cite{Fawcett04,Shannon69}, SLIC has five fitting parameters.  However, if parameterized to reproduce explicit-solvent simulations, the model has only three fitting parameters, which describe the nonlinear susceptibility in the solvation layer~\cite{Bardhan14_asym,Bardhan15_PIERS,Molavi16_JCTC}.   Nevertheless, the model gives excellent results: the RMS error is 1.3~kJ/mol for cations and ~2.5~kJ/mol for anions, in the 9 mixtures for which we have experimental data. Considering the model's simplicity, lack of chemical detail, and robustness to different solvents, this accuracy is surprising, because it suggests that specific chemical interactions such as hydrogen bonds need not be explicitly included for predictive accuracy.  This work addresses only monovalent ions, because polyvalent ions induce dielectric saturation in the first shell, introducing an additional nonlinearity between the first and second shells~\cite{Jayaram89,Alper90,Sandberg02,Kornyshev97,Marcus91}. Ongoing work aims to extend SLIC to model polarization saturation around highly charged solutes.

The paper is organized as follows. The following section presents the SLIC model for the electrostatic component of molecular solvation free energies.  Section~\ref{sec:thermodynamics} then addresses the application of SLIC to ion solvation thermodynamics in multiple polar solvents, and in Section~\ref{sec:mixtures} we study ion solvation free energies in mixtures.  Section~\ref{sec:discussion} concludes the paper with a discussion of open questions, limitations, and areas for future work.

\section{Theory}

Our model assumes that the solvation free energy can be decomposed as $\Delta G^{solv} = \Delta G^{np} + \Delta G^{es}$, where $\Delta G^{np}$ represents the nonpolar free energy associated with growing a completely uncharged solute cavity into the solvent, and $\Delta G^{es}$ represents the free energy of creating the solute charge distribution~\cite{Roux99}.  Because we are studying monovalent Born ions, we follow the typical convention and assume $\Delta G^{np}$ is negligible, i.e. in this paper we consider only the electrostatic solvation free energy, and assume $\Delta G^{solv} = \Delta G^{es}$.

In the standard (macroscopic) dielectric continuum model for $\Delta G^{es}$, the solute is modeled as a dielectric medium with relative permittivity $\epsilon_{in}$ that contains $N_q$ charges, usually at the atom centers (the $i$th charge is $q_i$ and located at $\bm{r}_i$), and the solute potential satisfies the Poisson equation. The solvent exterior is modeled as an infinite homogeneous bulk dielectric with relative permittivity  $\epsilon_{out}$, and in the absence of mobile charges (that is, in non-ionic solution), the solvent potential satisfies the Laplace equation. It is assumed that  $\phi_{out}\rightarrow 0$ as $|\bm{r}|\rightarrow \infty$, and that the normal flux across the dielectric interface (denoted $S$) is given by the standard Maxwell interface condition
\begin{align}
\epsilon_{in}\frac{\partial \phi_{in}}{\partial n}(\bm{r}_{S-})&=\epsilon_{out}\frac{\partial \phi_{out}}{\partial n}(\bm{r}_{S+}),\label{eq:SMBC}
\end{align}
where $\partial/\partial n$ denotes the outward normal derivative, $\bm{r}_{S-}$ is a point just inside the dielectric boundary $S$, and $\bm{r}_{S+}$ is a point just outside. Solving this problem using  finite difference methods or boundary integral methods, we obtain the reaction potential $\phi_{reaction}$ which arises due to the different permittivities. We write the electrostatic component of the solvation free energy as
\mbox{$\Delta G^{es} = \Delta G_{reaction} = \frac{1}{2} \sum_{i=1}^{N_q} q_i \phi_{reaction}(\bm{r}_i)$}
where $\phi_{reaction}(\bm{r})$ is the reaction potential field

\begin{figure}[H]
   \begin{center}
     \includegraphics[width=5in]{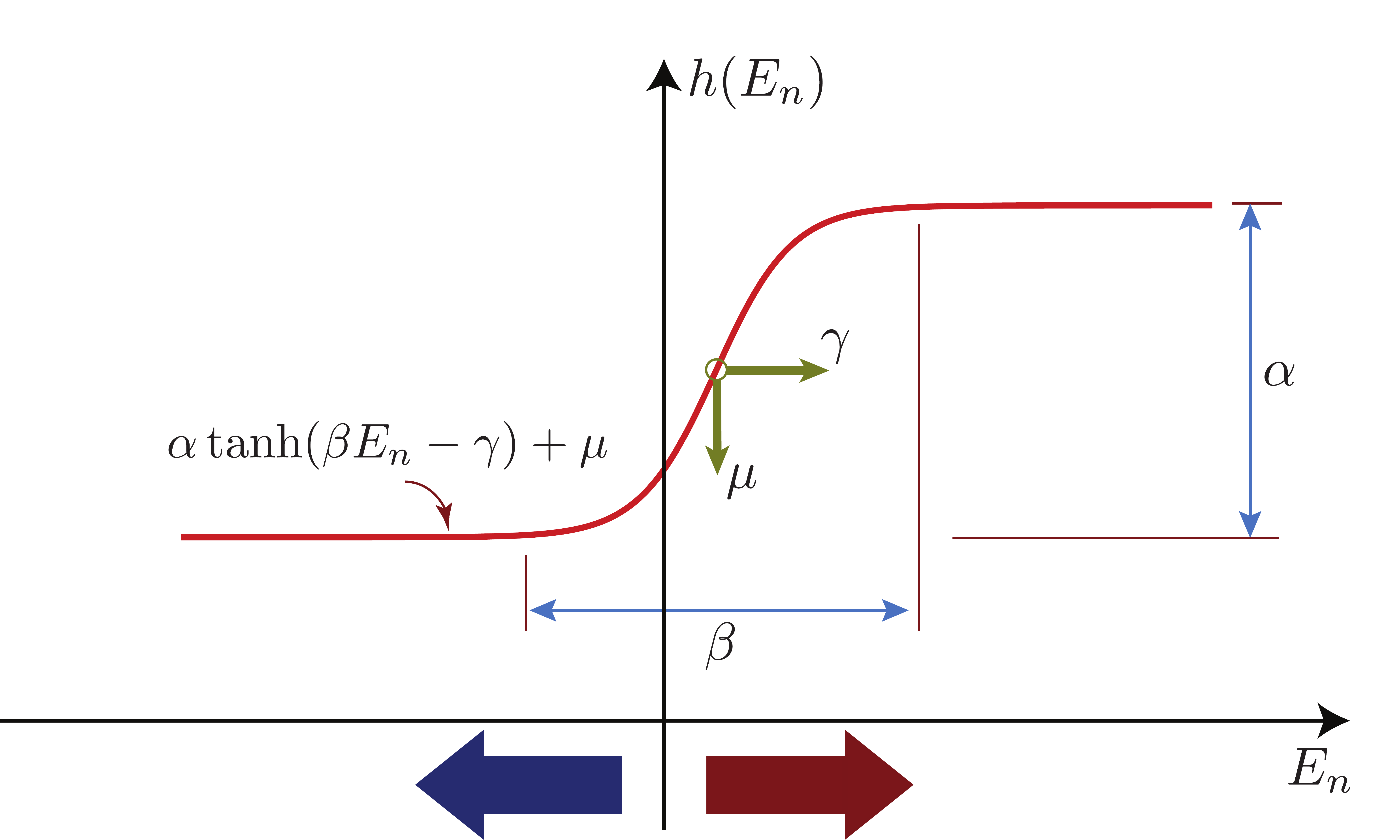}
   \end{center}
 \caption{Schematic of the SLIC perturbation to the standard dielectric interface condition, and the different model parameters.}
 \label{fig:tanh}
\end{figure}

In the SLIC model, by contrast, $\Delta G^{es}$ is defined to be the sum of two terms:
\begin{align}
\Delta G^{es}=\Delta G_{static}+\Delta G_{reaction}.\label{eq:separate-reaction-static}
\end{align}
The first term in Eq.~\ref{eq:separate-reaction-static} captures the component of the charging free energy that arises due to the interfacial potential field $\phi_{static}(\bm{r})$ created by solvent structure around a completely uncharged solute (i.e., an empty cavity with the solute shape) \cite{Cerutti07,Bardhan12_asymmetry}.  This term has been omitted in most previous dielectric continuum models, which leads to apparent deviations at very low charge densities~\cite{Nina97,Bardhan12_asymmetry,Lin14}.  In this work, we assume the static potential field $\phi_{static}$ is constant everywhere inside the solute; validation and justification for this approximation can be found in \cite{Cerutti07,Bardhan12_asymmetry,Bardhan14_asym}.
The second term in Eq.~\ref{eq:separate-reaction-static} is the familiar polarization energy associated with solvent polarization in response to the solute charge distribution. However, in contrast to the standard dielectric model, we have replaced the dielectric interface condition of Eq.~\ref{eq:SMBC} with the solvation-layer interface condition (SLIC)~\cite{Bardhan14_asym,Bardhan15_PIERS}:
\begin{align}
  \left(\epsilon_{in}-\Delta\epsilon \; h\left(E_n(\bm{r}_{S-})\right)\right)\frac{\partial \phi_{in}}{\partial n}(\bm{r}_{S-})=
  \left(\epsilon_{out}-\Delta\epsilon \; h\left(E_n(\bm{r}_{S-})\right)\right)\frac{\partial \phi_{out}}{\partial n}(\bm{r}_{S+})
\label{eq:slic0}
\end{align}
where $\Delta\epsilon=\epsilon_{out}-\epsilon_{in}$ and $E_n(\bm{r}_{S-})$ is the normal electric field at $\bm{r}_{S-}$ (note that the electric field just outside the surface does not explicitly enter into the interface condition). Notice that this change makes the induced surface charge sensitive to the local electric field, and in particular changes the response to positive and negative fields, matching our intuition about asymmetric solvation by water molecules. The perturbation $h(E_n)$ is 
\begin{align}
h(E_n)=\alpha \tanh(\beta E_n -\gamma) +\mu.
\label{eq:hfunc}
\end{align}
Figure~\ref{fig:tanh} is a schematic plot of this perturbation.  In this function, $\alpha$ dictates the magnitude of the deviation between suppressed response and enhanced response; $\beta$ determines the change in electric field necessary to transition solvation-layer response between modes; $\gamma$ determines the critical electric field where the transition is centered; and $\mu$ determines where the suppressed response and enhanced response are situated with respect to bulk response. It is important to note that the system responds linearly in regions where the derivative of $h$ is zero.  Therefore, as the width of the transition approaches zero, the system obeys two different regimes of linear response depending on the local field~\cite{Bardhan12_asymmetry}.  A small but finite transition region allows the model to reproduce observed nonlinearities at low field strengths, which have been noted to arise due to transition of solvent dipole orientations~\cite{BJBerne1}.  However, for charged or highly polar compounds, this transition region's energetic contribution to solvation is quite small~\cite{Molavi16}. In particular, because the actual region of nonlinear response happens in a very narrow region around $E_n=0$~\cite{Bardhan14_asym,Bardhan15_PIERS,Molavi16}, the change in potential due to a change in solute charge is essentially linear for any finite charge, so the polarization component of the electrostatic solvation free energy can be approximated using the usual expression $\Delta G_{reaction}=\frac{1}{2}\sum q_i \phi_{reaction}(\bm{r}_i)$.

\section{Predicting Solvation Thermodynamics}\label{sec:thermodynamics}

To test whether the SLIC dielectric continuum model can reproduce solvation thermodynamics, and to assess the effects of the static potential on prediction accuracy, we calculated ion solvation free energies, entropies, and heat capacities in nine polar solvents: water (abbreviated W), methanol (MeOH), ethanol (EtOH), formamide (F), acetonitrile (AN), dimethylformamide (DMF), dimethyl sulfoxide (DMSO), nitromethane (NM), and propylene carbonate (PC). The test set was composed of the monovalent Born ions Li$^+$, Na$^+$, K$^+$, Rb$^+$, Cs$^+$, Cl$^-$, Br$^-$, and I$^-$; however, we could not use Rb$^+$ for MeOH, EtOH, F, DMSO, or NM due to a lack of experimental data. To parameterize the model and its temperature dependence, we used experimental solvation free energies at multiple temperatures; solvation free energy changes due to temperature were calculated using experimental solvation entropies and heat capacities~\cite{Chamberlin08}.  For each solvent/temperature pair, we parameterized the model once with $\phi_{static}$ set to zero, and once with it allowed to vary.    Note that ion radii were taken to be widely used values~\cite{Fawcett92} without any further adjustment.  Thus, for each solvent/temperature parameterization the fitting was overconstrained, having more data points (8 or 9, see below) than model parameters (4 or 5, depending on the use of $\phi_{static})$.  Other relevant details for the solvents can be found in the supporting information. 

\begin{figure}
  \includegraphics[width=2.5in]{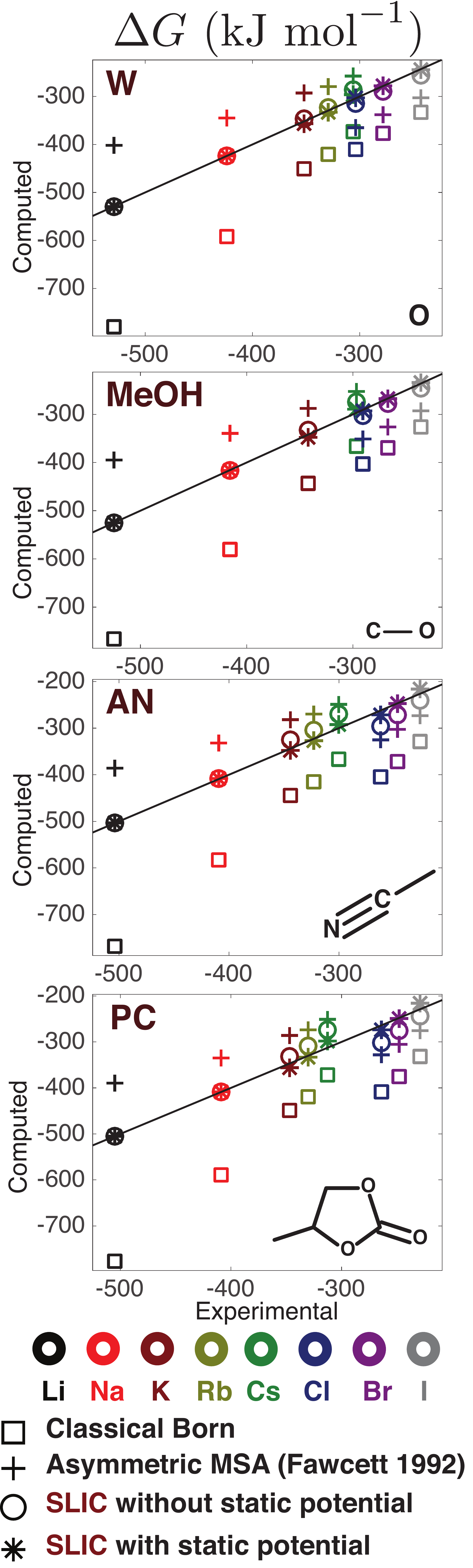}
  \caption{Solvation free energies calculated by the classical Born dielectric model, the asymmetric MSA~\cite{Fawcett92}, and the SLIC model with and without the static potential. The solid lines represent perfect agreement between experiment and theory.}
  \label{fig:dg}
\end{figure}

\begin{figure}
  \includegraphics[width=2.5in]{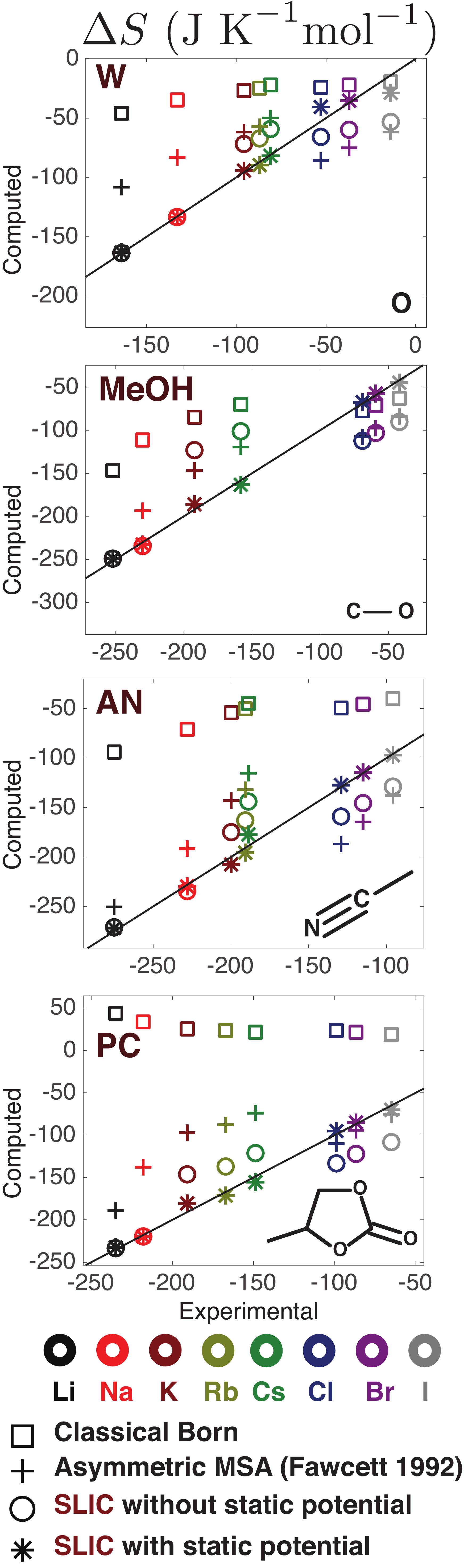}
  \caption{Entropies calculated by the classical Born dielectric model, the asymmetric MSA~\cite{Fawcett92}, and the SLIC model with and without the static potential. The solid lines represent perfect agreement between experiment and theory.}
  \label{fig:ds}
\end{figure}

Figures~\ref{fig:dg} and ~\ref{fig:ds} contains plots of SLIC predictions of ion solvation free energies and entropies at 25~C, along with predictions from standard Born theory and the asymmetric MSA theory~\cite{Fawcett92} for four solvents (W, MeOH, AN, and PC). The solvation free energies and entropy plots for other solvents are available in the supporting information. Both SLIC models are substantially more accurate than the existing models. It is also clear that the SLIC model \textit{with} the static potential is much more accurate than the one that omits it, especially for entropies (as well as free energies in F, AN, DMF, DMSO, NM, and PC).  Interestingly, for anions, the SLIC model predicts exaggerated entropy differences compared to experiment; this is particularly noticeable for W, NM, and DMF. In addition, the cation entropies are generally more accurate than the anion entropies, but larger cations in AN are an exception.  More detailed studies using explicit-solvent molecular dynamics are in progress.

\begin{figure}
  \includegraphics[width=2.5in]{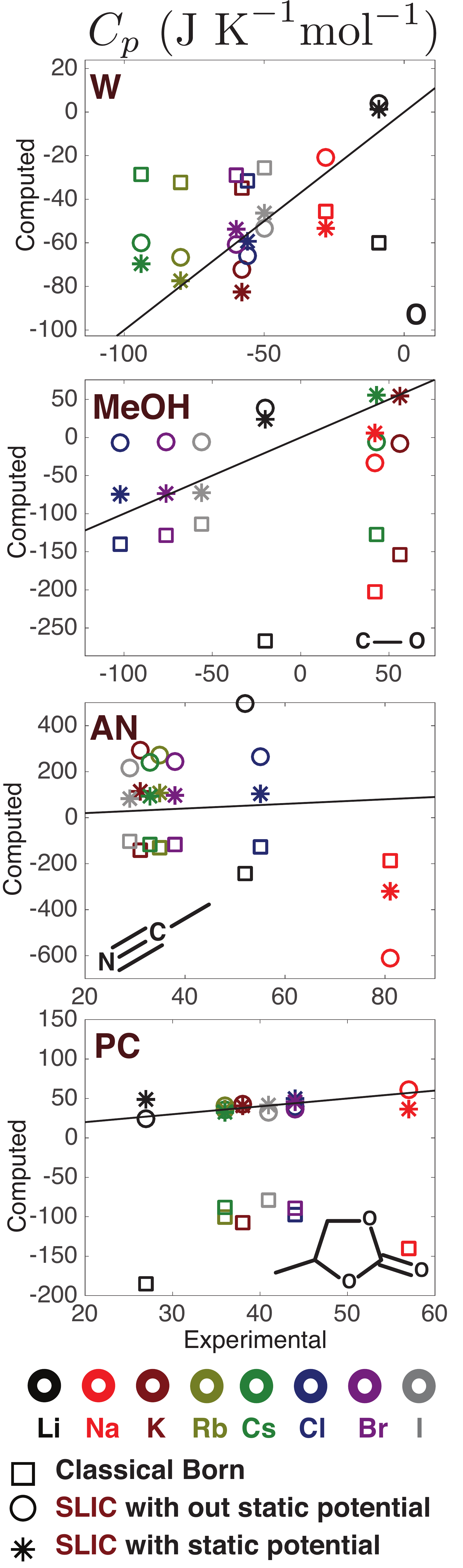}
  \caption{Heat capacities predicted by the classical Born theory and by the SLIC model with and without the static potential.  The solid lines represent perfect agreement between experiment and theory.}
  \label{fig:cp}
\end{figure}

Figure~\ref{fig:cp} contains plots of calculated heat capacities in  W, MeOH, AN, and PC compared to experimental data. Plots of the calculated heat capacities for other solvents can be found in the supporting information. Because heat capacities are related to the second derivative of the free energy, it is unsurprising that the correlations are weaker than for energies and entropies. As expected, the classical Born model is incapable of calculating heat capacities accurately~\cite{Vath99}. In our model, inaccuracies are particularly notable for anions, which may be related to their greater degree of charge transfer~\cite{Zhao10,Rogers10}.  The influence of the static potential on heat capacities is especially notable, providing an important offset to improve agreement with experiment in almost all cases.  These results suggest that the static potential (an intrinsic property of the solvent and only weakly dependent on the shape of the uncharged solute) has a substantial effect on solutes' heat capacities, but that in a given solvent, differences in heat capacities between molecules are governed by more detailed physics.  In addition, we observe that small cations are problematic, which is not surprising because their high charge density leads to dielectric saturation, meaning that discrete solvent structure becomes increasingly important.

Together, Figures~\ref{fig:dg},~\ref{fig:ds}, and~\ref{fig:cp} along with the corresponding figures in the supporting information indicate the importance of including the static potential in predicting solvation thermodynamics.  The results also suggest that SLIC works well for solvents of various structure, complexity, and size, even though solvent structural details are not addressed explicitly. A table containing the values of each SLIC parameter at $T=25^{\circ}$C, and their derivatives with respect to temperature, is available in Supporting Information. Previous work has shown that SLIC works well for polyatomic solutes such as biomolecules in water~\cite{Bardhan14_asym}, and future work will address such solutes in larger, more complex solvents such as PC.  The present results do, however, explain previously noted questions, such as why $\Delta S$ does not have a straightforward dependence on ion radius~\cite{Salomon70}: namely, the interface potential is largely independent of radius (though it does exhibit some variation~\cite{Ashbaugh00}).  Our model assumes $\phi_{static}$ is  independent of solute shape, yet predicts these quantities accurately, which suggests that its variation with size does not play a major role in ion solvation thermodynamics. We reiterate that the present model does not address second-shell effects~\cite{Bardhan15_PIERS}, which is why the ions considered are only monovalent. Polyvalent ions that saturate the first shell~\cite{Rashin85_Honig} will be studied in future work.

Having established the SLIC models' accuracy, we next consider interpretation of the model parameters.  Because the SLIC model without $\phi_{static}$ exhibits demonstrably poorer accuracy, we study only the model that includes it. Because we used Shannon and Prewitt radii, the parameter values cannot be interpreted directly. That is, explicit-solvent simulations and a SLIC model based on MD radii are needed to provide a consistent model comparison and offer atomistic insights into the model's treatment of solvation-layer response.  However, it is worth noting common features of the temperature-dependent response.  For all solvents, the width of the transition region in Figure~\ref{fig:tanh} (as captured by $1/\beta$) is increasing with temperature, which can be interpreted in terms of increased thermal motion leading to more gradual transition, as a function of the local electric field.  Results for $\alpha(T)$, the magnitude of the response asymmetry (between enhanced and suppressed response) are not consistent: W, MeOH, AN, NM, and PC exhibit increases in response asymmetry with temperature (positive $\frac{\partial \alpha}{\partial T}$), while the others exhibit decreasing trends.  The centering parameters $\mu$ and $\gamma$ increase with temperature for all solvents, but the significance of these variations are not clear.  The static potentials for all solvents are negative, and increasing (becoming less negative) with temperature, but more detailed simulation will be required to establish the relationship to microscopic phenomena.

We may also consider these solvents from the perspective of being protic or aprotic. The protic solvents water, MeOH, and EtOH exhibit positive correlations between dielectric constant and  $\alpha$ and $\beta$, but negative correlations between the dielectric constant and $\gamma$ and $\mu$. However, formamide does not follow this trend. For aprotic solvents, the parameters do not exhibit any obvious dependency on the dielectric constant or the solvent radius (supporting information). Future work will address these relationships in more detail, and particularly focus on  the extent to which solvation entropies and heat capacities are in fact related to the parameters' temperature dependence.

Table~\ref{tbl:prediction} contains calculations for solvation thermodynamics where experimental data are not yet available or were not used in parameterization.  These cases include F$^-$ in all solvents, and Rb$^+$ in MeOH, EtOH, F, DMSO, and NM. Because solvation of fluoride ions in nonaqueous solvents has received limited attention~\cite{Hefter89}, we did not use this ion for parameterizing SLIC in any of the solvents, even in water, where it has been studied~\cite{Fawcett04, Marcus96}. Table~\ref{tbl:prediction} also includes available experimental measurements.  Again, it can be seen that the model predicts free energies and entropies accurately, and is qualitatively reasonable for heat capacities (especially compared to other models). 
\begin{table}
  \resizebox{.8\textwidth}{!}{%
  \begin{tabular}{ c |c |c |c |c}
   Solvent & Ion &$\Delta G$   (kJ mol$^{-1}$) &$\Delta S$   (J K$^{-1}$ mol$^{-1}$) & $C_p$   (J K$^{-1}$ mol$^{-1}$) \\
  \hline
  W					& F$^-$     	&		-430 (-429)\cite{Fawcett04}		&		 -67 (-115)\cite{Fawcett04} 		&	-86 (-45)\cite{Marcus96} 			\\
 \hlineB{3}
 \multirow{2}{*}{MeOH} 	& Rb$^+$  	&		-326(-319)						&		 -178 (-175)		&	55				\\
					\cline{2-5}
					& F$^-$     	&		-415							&		 -116 			&	-79 (-131)\cite{Marcus96}			\\
 \hlineB{3}
\multirow{2}{*}{EtOH} 	& Rb$^+$    	&		-319 (-313)					&		 -197 (-187) 		&	128 				\\
					\cline{2-5}
					& F$^-$	    	&		-405				&		 -145 			&	-153 (-194)					\\				
 \hlineB{3}
  \multirow{2}{*}{F} 		& Rb$^+$    	&		-340 (-334)		&		 -135 (-130) 	&	27								\\
					\cline{2-5}
					& F$^-$	    	&		-418				&		 -128			&	36 (28)\cite{chen98_Hefter}				\\				
 \hlineB{3}
  AN					& F$^-$	    	&		-390				&		 -192			&	147 					\\	
 \hlineB{3}
   DMF				& F$^-$	    	&		-389				&		 -230 		&	105					\\	
 \hlineB{3}
  \multirow{2}{*}{DMSO} 	& Rb$^+$    	&		-348 (-339)		&		 -151 (-180) 	&	32  					\\
					\cline{2-5}
					& F$^-$	    	&		-400				&		 -160		&	186(60)\cite{Marcus96}				\\				
 \hlineB{3}
\multirow{2}{*}{NM}	 	& Rb$^+$    	&		-324 (-318)		&		 -186 (-183) 	&	19 					\\
					\cline{2-5}
					& F$^-$	    	&		-391				&		 -182			&	95(71)\cite{Marcus96}				\\				
 \hlineB{3}
 PC					& F$^-$	    	&		-394				&		 -149 		&	67					\\

  \end{tabular}}
\caption{Prediction of Gibbs free energy, entropy and heat capacity in the model with $\phi_{static}$. Values in parentheses are experimental values when available.}
  \label{tbl:prediction}
\end{table}

\section{Predicting Solvation in Mixtures}\label{sec:mixtures}
We parameterized concentration-dependent SLIC models for ion solvation in 9 water--co-solvent mixtures.  The  co-solvents were acetone (AC), acetonitrile (AN), dioxane (Diox), dimethyl ether (DME), dimethylformamide (DMF), dimethyl sulfoxide (DMSO), ethanol (EtOH), methanol (MeOH), and urea.   We obtained ion solvation free energies in each mixture by adding tabulated transfer free energies~\cite{Wells73-2,Wells73,Wells84,Wells81,Groves85,Groves85-2,Wells92,Wells78,Sidahmed84} to experimental ion solvation free energies in water~\cite{Fawcett04}.  Mixture dielectric constants were taken to be experimental values~\cite{Suresh02,Markarian09,Akerlof32,Roy02,Wyman33,Douheret88,Akerlof36,Kumbharkhane93}.  The experimental transfer free energies included the monovalent Born ions Li, Na, K, Rb, Cs, Cl, Br, and I, though transfer free energies were not available for every ion in every co-solvent.  Each SLIC parameter was modeled as varying quadratically (for example, $\alpha(c) = \alpha_0 + \alpha_1 c + \alpha_2 c^2$) where the co-solvent weight/weight concentration $c$ between $0$, meaning pure water, and a maximum of 1, pure co-solvent.  However, transfer free energies from pure water to pure co-solvent were not available. Thus, for each solvent, the 5 SLIC dependent parameters led to a fitting of 15 parameters over all experimental data associated with that co-solvent mixture, regardless of concentration. For each optimization, every solvation free energy was weighted equally in the optimization problem, and every co-solvent had at least 36 measured transfer energies.  Therefore, each optimization problem was well posed.  Again, no ion radii were fit during this work: the Shannon--Prewitt radii were used unchanged~\cite{Shannon69,Fawcett04}.  The optimization problems were unconstrained, and for initial guesses we used coefficients obtained by polynomial fitting from parameterizations at individual mixture concentrations.  We verified the model consistency by using the optimized SLIC models of different mixtures to predict solvation free energies in neat water (Supporting Information).  MATLAB's non-linear least squares function was used for optimization.  

Table~\ref{table:RMSerrors} contains the root-mean-square (RMS) errors for the SLIC model associated with each co-solvent, tabulated separately for cations and anions.  Errors are given for both the absolute solvation free energies and for the transfer free energies from neat water to a given mixture.  The model achieves high accuracy, with RMS errors for solvation free energies less than 7~kJ/mol.  However, differences can be observed in that the cation predictions are somewhat less accurate than predictions for anions.  Transfer free energies (measured as the solvation free energy difference between the model at 0\% co-solvent and the model at finite co-solvent weight fraction) are highly accurate, with both cations and anions achieving RMS errors of less than 2.5~kJ/mol, though cation transfer free energies are more accurate than those for anions. 

\begin{table}[H]
\begin{center}
\begin{tabular}{|c|c|c|c|c|}
\hline
& \multicolumn{2}{|c|}{ $\Delta G^{solv}$} & \multicolumn{2}{|c|}{$\Delta G^{tr}$} \\\hline
\textbf{Solvent} & Cations & Anions & Cations & Anions
\\\hline
\textbf{AC}&6.64&1.10&0.13&1.38\\
\textbf{AN}&2.25&1.27&0.86&0.95\\
\textbf{Diox}&2.71&1.95&1.09&1.81\\
\textbf{DME}&5.07&1.14&0.42&1.50\\
\textbf{DMF}&4.62&0.75&1.29&1.07\\
\textbf{DMSO}&2.22&2.58&0.47&1.71\\
\textbf{EtOH}&4.45&2.41&0.86&2.46\\
\textbf{MeOH}&2.24&1.50&0.41&0.54\\
\textbf{Urea}&2.48&1.63&0.72&0.53\\\hline
\end{tabular}
\caption{RMS errors, in kJ/mol, for $\Delta G^{solv}$ and $\Delta G^{tr}$, computed separately for cations and anions.} \label{table:RMSerrors}
\end{center}
\end{table}

\begin{figure}
  \includegraphics[width=2.7in]{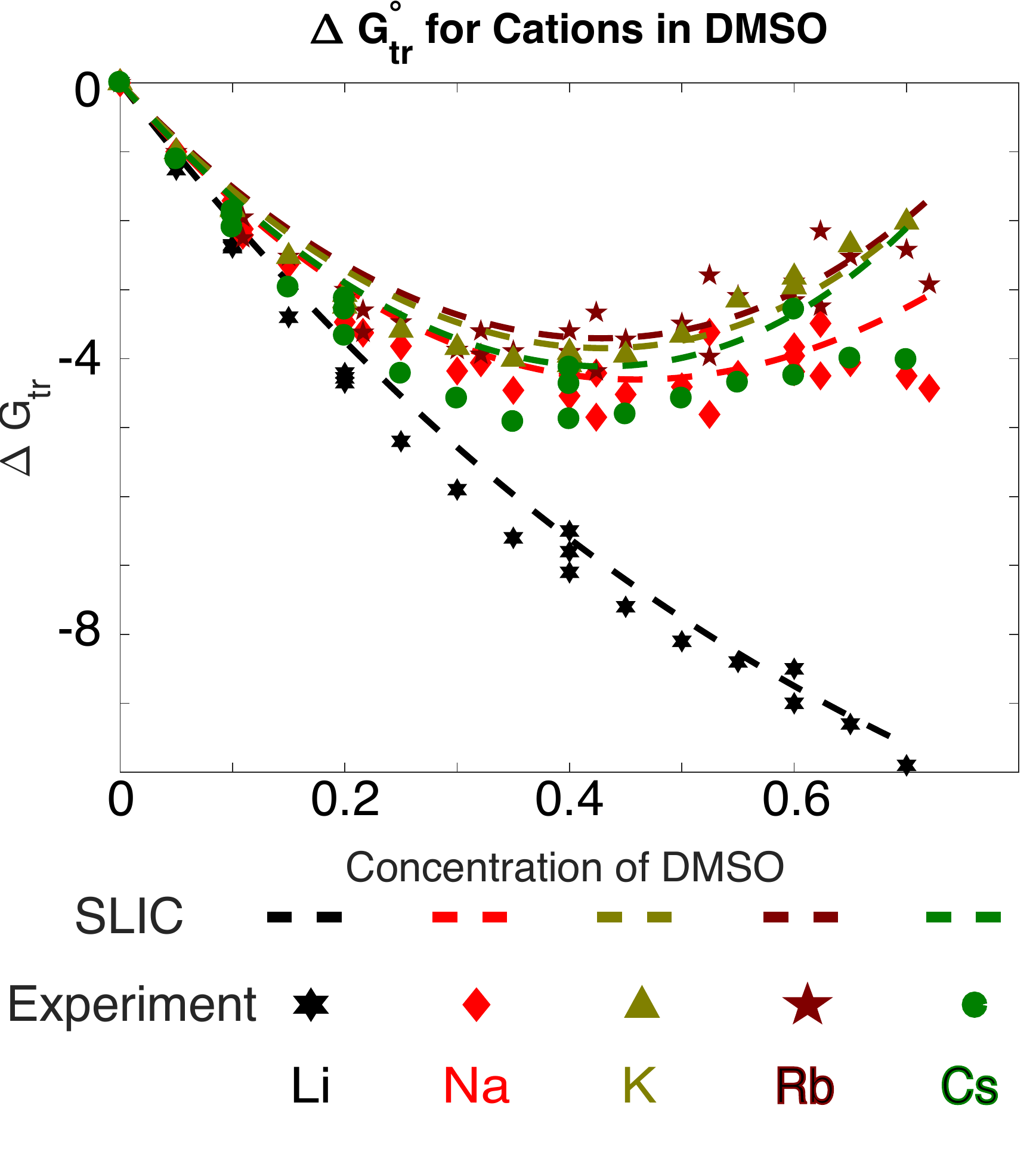}
  \caption{Transfer free energies, in kJ/mol, for cations into water-DMSO mixtures.}
  \label{fig:cations-dmso}
\end{figure}

\begin{figure}
  \includegraphics[width=2.7in]{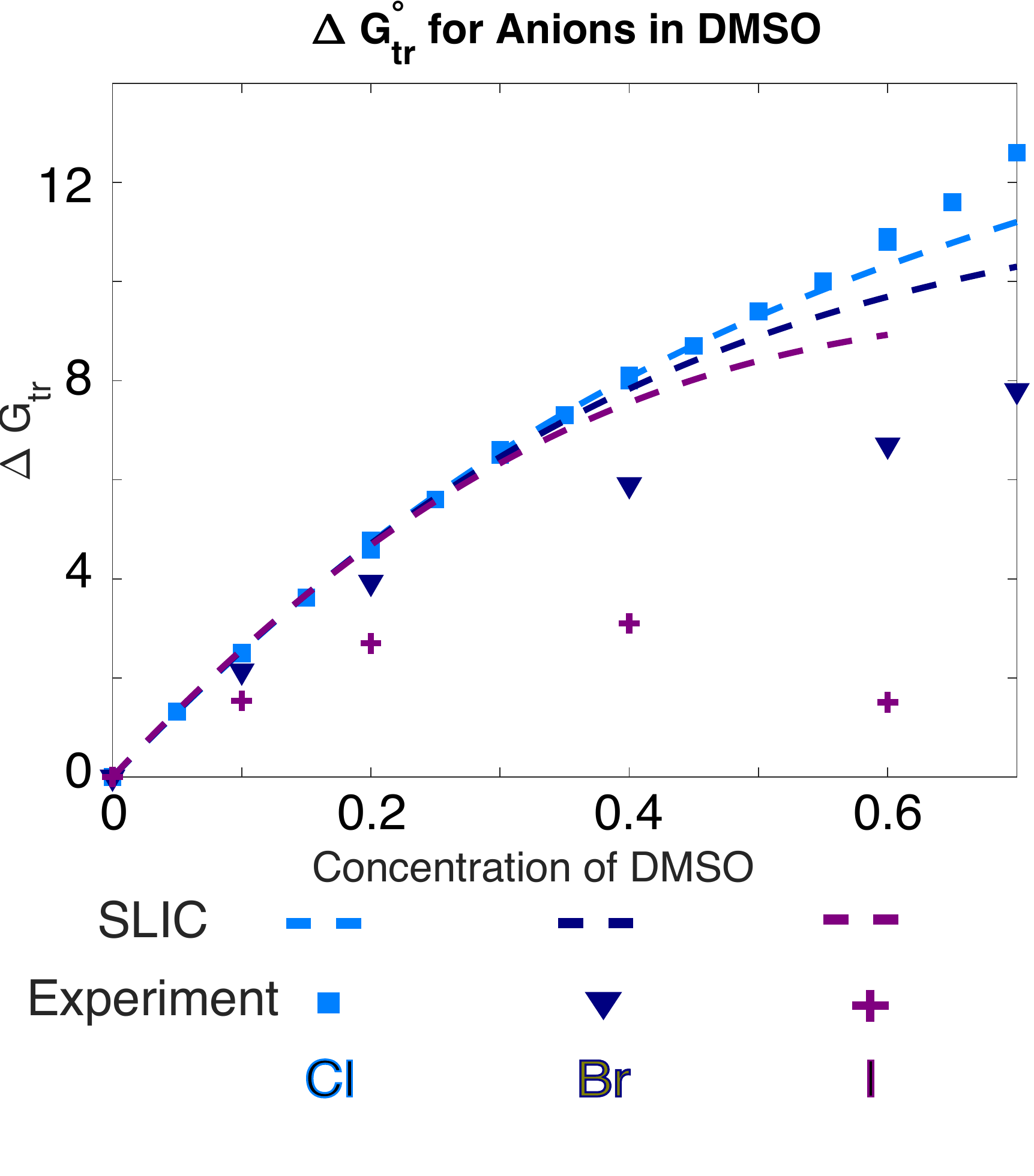}
  \caption{Transfer free energies, in kJ/mol, for anions into water-DMSO mixtures.}
  \label{fig:anions-dmso}
\end{figure}

Figures~\ref{fig:cations-dmso} and~\ref{fig:anions-dmso} are plots of the cation transfer free energies and anion transfer free energies, respectively, into mixtures of water and DMSO; these are representative of the results for all solvents.  The Supporting Information contains individual plots for the transfer free energy profile for each Born ion in each co-solvent mixture, compared to both experiment and the prediction of continuum Born theory.  For the Born model we held the Shannon-Prewitt radii fixed but changed the dielectric constant according to experiment.  The cation transfer free energy profiles are well reproduced in our theory;  cesium, the largest, is underpredicted by a small but consistent amount.  For anions, the experimental profiles exhibit a wide variance as the concentration increases (Figure~\ref{fig:anions-dmso}); these results are observable to a lesser extent for other mixtures, including ethanol, DMF, and dioxane.  With regard to relative transfer free energies between cations, our model reproduces experimental orderings reasonably well over the concentration range for which experiments are available, with the exception of potassium. For anions, however, the SLIC differences are underpredicted compared to the experimental measurements.  The complete set of transfer free energy profiles are available in the Supporting Information, offering additional evidence of the SLIC model's accuracy.  First, in DMSO as well as in ethanol, urea, DMF, DME, and dioxane, SLIC reproduces cations' concave-up transfer free energy profiles and concave-down profiles for anions.  Furthermore, in water-methanol mixtures, the cation profiles are concave down and the anions concave up, and SLIC reproduces this difference (although predictions for potassium exhibit poorer accuracy).  Second, for acetone and acetonitrile, the anion transfer free energies vary essentially linearly with concentration, which our model also reproduces.  Third, transfer free energies for cations in ethanol mixtures exhibit an inflection point, and our model reproduces the overall profiles accurately, though not the change in curvature.

We also show in the Supporting Information that with fixed radii, the classical Born model, which uses purely macroscopic dielectric notions, is unable to reproduce even qualitative features, because the only varying parameter is the dielectric constant.  For example, the transfer free energy profiles often have significant curvature and a local maximum or minimum, whereas the Born-model profiles are monotonic (see particularly the results for dioxane).  To construct an accurate Born model, each ion's radius must be parameterized at each co-solvent concentration; one observes a non-monotonic variation in radius that can be as large as 0.1~\AA~(Supporting Information). As a blind prediction to test the SLIC model, the Supporting Information also includes predictions for ion transfer free energies in co-solvent mixtures for which we did not find any reference data.  These predictions included fluorine for all co-solvents, as well as lithium and sodium in acetone and ethanol.  For water-ethanol mixtures, the lithium and sodium transfer free energy profiles are very similar to the other cations' profiles.  In contrast, the predictions for acetone are quite different for larger cations, which suggests that such experiments or atomistic simulations would offer a stringent test of our model.


%

\section{Discussion}\label{sec:discussion}

We have established that a dielectric continuum solvent model can accurately reproduce ion solvation thermodynamics in a variety of polar solvents and solvation free energies in mixtures, provided that (1) the usual macroscopic dielectric interface condition is replaced with a solvation-layer interface condition (SLIC), and (2) proper account is taken for the interface potential, which we have  termed a static potential in order to highlight its microscopic character~\cite{Cerutti07}.  Importantly, the SLIC model achieves this accuracy via a simple physical picture rather than a chemical one---that is, SLIC does not account explicitly for solvent chemical properties such as hydrogen bonding propensities or detailed solvent structure.  Instead, the model treats in essence the re-orientation response of asymmetric dipoles in the solvation layer. Overall, however, SLIC works very well for both protic and aprotic solvents, as well as for solvents of varying size and structure; other work has established its accuracy for predicting solvation free energies of polyatomic solutes~\cite{Bardhan14_asym,Bardhan15_PIERS}.  Our assessment of transfer free energies in mixtures offers encouraging signs for the model's robustness.  For mixtures, our results were obtained via global optimization (over all concentrations for a given co-solvent) with parameters varying quadratically as a function of concentration.  However, even linear dependencies work reasonably well for most solvents, despite the reduced number of fitting parameters (Supporting Information).  Furthermore, SLIC models implemented to match explicit-solvent MD need even fewer parameters~\cite{Bardhan14_asym}.  SLIC predicts, with semi-quantitative accuracy, the experimental free energies of transfer over a wide range of concentrations, even when the dependencies have different trends over the Born ions.  The accuracy and generality suggest that first-shell solvent response, as captured via a surface-charge representation, suffices to explain a large component of changes in solvation over substantial changes in solvent composition.  To put our present model to a stringent test, we have predicted solvation thermodynamics and transfer free energies for cases in which we have no experimental data (primarily fluorine, as well as lithium and sodium; see Supporting Information).  Our calculations of entropies and heat capacities also indicate the importance of separating the static potential from the nonlinear polarization response.

In work on polyatomic solutes including amino acids, we have shown that the model does not require atom radii to be adjusted for atomic charge~\cite{Bardhan14_asym}, which differs from numerous suggestions and parameterizations. The present work shows that the SLIC continuum model is highly accurate even when the solute atom radii are independent of solvent composition and temperature~\cite{Elcock97,Nina97}.  In contrast to models which parameterize many radii (making model comparison challenging) what changes in SLIC is the interface condition, and optionally the static potential.  In our view, this is a more meaningful adjustment because the system changes involve the solvent and the solvent--solute interactions, not in the solute itself.  We note that this viewpoint is implicit in the MSA model for Born ion solvation~\cite{Fawcett04}.  Our model's rather surprising accuracy provides further support that temperature-dependent changes in the average charge structure of the solvation layer, rather than specific chemical interactions, are responsible for ion solvation thermodynamics.  We have predicted solvation entropies with high accuracy and heat capacities with only reasonable accuracy, but this lower accuracy is not surprising given that heat capacities are second-derivative quantities, and more chemical detail is likely to be needed for these predictions.

For mixtures, straightforward calculations illustrate a clear weakness of classical dielectric models: Born radii fit to experimental results must vary non-monotonically with co-solvent concentration.  This firmly establishes the notion that in traditional Poisson models, the atom radii must be considered as free (adjustable) parameters~\cite{Papazyan97}.  In contrast, all of our calculations here used the standard Shannon--Prewitt radii~\cite{Shannon69}; when SLIC is parameterized against explicit-solvent MD simulations, the resulting model is accurate using standard MD Lennard-Jones radii with only a uniform scaling~\cite{Bardhan14_asym,Molavi16_JCTC}.  In this respect, our model has dozens of fewer fitting parameters than traditional continuum electrostatic models, where radii must be fit for each atom type or for many groups of similar type. It is worth emphasizing that recent models of charge-hydration asymmetry are similarly able to reproduce wide sets of experimental data using fewer radii fitting parameters~\cite{Corbeil10,Mukhopadhyay14} than classical continuum models.

A question of significant interest is how to reconcile the solvation-layer response picture of SLIC with the significant literature on the role of solvent fluctuations~\cite{Hummer95,Hummer97,Martin08,Sarupria09}, which our model does not include.  For example, can the solvent fluctuation density field be decomposed into terms related to the static and reaction fields? It is also interesting to consider the relationship of our approach to the local molecular field (LMF) theory of Weeks et al., which approximates the exact Yvon--Born--Green hierarchy~\cite{Rodgers08,Remsing16}.  Both models determine the electrostatic potential field inside the solute, and could be compared in fine detail.  In ongoing work we are assessing the SLIC model's capacity to predict the stabilities of cation-anion contact pairs in solution, and the impact of including SLIC in the polarizable continuum model (PCM)~\cite{Miertus81,Mennucci10}.  Compared to existing implicit-solvent models for mixtures, SLIC differs in three primary ways. First, numerous models have been proposed for specific co-solvents, but to our knowledge only COSMO-RS and RISM-based models have been demonstrated on the large number of co-solvents as we have shown here.  We have also shown that our model reproduces experimental trends in transfer free energies with high accuracy; that is, our model captures dependencies on concentration, in addition to being accurate at specific co-solvent concentrations.  Second, SLIC has already been shown to work very well for polyatomic solutes with complex geometries; in contrast, many existing models have focused only on spherical ions or spherical nanoparticles.  The exceptions here again are RISM-based models and COSMO-RS.  Third, SLIC represents a remarkably small modification of traditional Poisson--Boltzmann based dielectric models, and can be incorporated easily into the large number of finite-difference, finite-element, or boundary-element solvers~\cite{Bardhan15_PIERS,Molavi16_JCTC}.   

The model's simplicity comes with attendant limitations and open questions, and the tests presented here cover only a fraction of possible applications.  Results on mixtures suggest that accuracy tends to decrease at high co-solvent concentrations.  These deviations were surprising given the model's accuracy for the neat co-solvents.  Unfortunately, the neat co-solvent solvation free energies were inconsistent with the transfer free energies available to us, precluding their use as data points at 100\% concentration.  We hope that future experimental measurements or explicit-solvent simulations may provide insights into these errors.  Ongoing work aims to predict the solvation of polyatomic solutes in mixtures and to investigate whether SLIC can predict molecular solvation thermodynamics in mixtures as it can in neat solvents.  We have also not yet tested the model on mixtures of polar and non-polar solvents.  There exist several implicit-solvent models for such mixtures~\cite{Basilevsky09,Basilevsky11,Bonincontro06}, and whether SLIC works for these solutions is not known.  Other current work extends our analysis here to a SLIC variant that can model dissolved ions in the solvent mixture using the linearized Poisson--Boltzmann equation~\cite{Molavi16_JCTC}.  This requires an additional nonlinear interface condition at the Stern (ion-exclusion) surface, and in mixtures the width of this ion-exclusion region will presumably depend on the co-solvent size and concentration.  We have distinguished the static potential field from the macroscopic notion of an interface potential. Because it arises from mean solvent structure around a solute, the field satisfies the Poisson equation; however, steric considerations mean that the static potential near the boundary is not actually uniform in a thin region at the surface of the solute (the first layer of solute atoms)~\cite{Cerutti07,Bardhan14_asym}.  Future work will investigate whether biological systems exploit this non-uniformity for molecular function, which may necessitate the development of a more sophisticated static potential model than the present assumption of a uniform field.  Future work will also investigate whether the functional form of the solvation-layer correction may be better fit to an error function than the present hyperbolic tangent~\cite{Sides99}. One additional open question is the whether SLIC can be applied to understand protein behavior in mixtures of water and osmolytes or denaturants.

Our development of SLIC originally only focused on solutes in water, and arose from a question that included theoretical, philosophical, and practical considerations: What would an accurate implicit solvent model look like if one did not specifically parameterize atom radii but simply used the values employed in MD? Theoretically, we were curious why an implicit-solvent model should need to use different atom radii depending on the sign of its charge.  For a monoatomic ion of a given chemical radius, its reaction potential can always be written in terms of an appropriate surface charge on a sphere of that radius, but the surface charge density might depend on the sign (and magnitude) of the charge. Philosophically, it seemed reasonable to consider that the solute atom did not change physically when embedded in a solvent of different composition or temperature, so the use of a state-dependent radius seemed like a way to correct deeper problems with the dielectric theory.  Practically, our development of SLIC arose from a simple motivation: the desire to avoid the need for extensive parameterization of radii with every new solvent theory.  A number of more proper justifications may be offered as well. First, there is an increasing interest from environmental and biotechnological research in the prediction of protein function at different temperatures. Second, both basic and applied biosciences research focuses on the effects of changing solution conditions such as the addition of co-solvents, or partition coefficients for transfer free energies between neat solvents~\cite{Helmer68,Lipinski97}.  Third, the costs and complexity of continuum-model parameterization and validation seem to be limiting the community's ability to use continuum models to address the massive chemical diversity associated with post-translational modifications of proteins. Fourth, there exist already a wide range of continuum solvers based on the PB theory, including large-scale parallel codes~\cite{Baker01}, codes coupled to MD~\cite{Cai10}, and many in quantum chemistry~\cite{Miertus81}.  The model's successes in this work and other recent studies motivate adapting some of these software packages for more challenging tests of the SLIC model.

\section*{Acknowledgments}
The authors thank F. C. Pickard IV, B. Brooks, M. Reuter, A. E. Ismail, P. Jungwirth, D. Green, and E. Boyden for valuable discussions. MGK was partially supported by the U.S. Department of Energy, Office of Science, Advanced Scientific Computing Research, under Contract DE-AC02-06CH11357, and the National Science Foundation under award numbers SI2-SSI 1450339 and  OCI-1147680. The work of JPB, SG, and AMR has been supported in part by the National Institute of General Medical Sciences (NIGMS) of the National Institutes of Health (NIH) under award number R21GM102642, and by a contract with Battelle. The content is solely the responsibility of the authors, and does not necessarily represent the official views of the National Institutes of Health.

\section*{Supporting Information}
The MATLAB source code for the solvation thermodynamics calculations can be accessed at: \url{https://bitbucket.org/bardhanlab/slic_solvation_thermodynamics}{ }. The MATLAB source code for the solvent mixture calculations can be accessed at \url{https://bitbucket.org/bardhanlab/si-slic-mixtures}{ }.  Supporting information for the thermodynamics calculations includes solvent details, SLIC parameters at $T=25^{\circ}$C and their derivatives with respect to temperature, and the full set of plots for solvation free energies, entropies, and heat capacities in all neat polar solvents.   Supporting information for the solvent mixtures calculations include (1) plots of all transfer free energy profiles for Born ions in all 9 co-solvent mixtures, compared to experiment and the classical Born model, under three types of parameterized SLIC models: quadratic concentration-dependence (discussed in this paper), a model with linear concentration-dependence (fewer fitting parameters), and a model with quadratic concentration-dependence where missing experimental data has been supplanted with interpolated results from polynomial fits to experiment; (2) validation of the co-solvent models' consistency by calculation of solvation free energies in neat water (i.e. at 0\% co-solvent); (3) RMS errors for the cations and anions in different co-solvent mixtures, for the three types of parameterized SLIC models; (4) concentration-dependent Born radii for ions in water-ethanol mixtures.   

\bibliographystyle{unsrt}
\bibliography{bardhan-lab}

\newpage



\section*{Supplementary material}

Supporting information for
"Predicting Solvation Free Energies in Mixtures of Polar Solvents Using a Solvation-Layer Interface Condition"
by
S. Goossens, A. Molavi Tabrizi, A. Mehdizadeh Rahimi, M. G. Knepley, and J. P. Bardhan

1. Plots of all transfer free energy profiles for Born ions in all 9 co-solvent mixtures, compared to experiment (data points) and Born theory (red curve).

2. Plots of transfer free energy profiles for all cations (left side) and all anions (right side), compared to experiment (data points).

3. Predictions of absolute ion solvation free energies in neat water, for each co-solvent mixture. Because mixtures were parameterized individually over the whole concentration range from 0% (neat water) to the highest co-solvent concentration data, these predictions for neat water represent a validation that the model produces a consistent parameterization.

4. Predicted transfer free energies for ions in co-solvent mixtures where no experimental data were available for the ion-cosolvent pair.

5. Born radii optimized to fit experimental data on transfer free energies (using the experimentally appropriate dielectric constant for each mixture).

6. RMS error table for the quadratically varying model (what is reported in the paper).

7. Plots of all transfer free energy profiles for Born ions in all 9 co-solvent mixtures using a LINEARLY varying set of SLIC parameters as a function of concentration, compared to experiment (data points) and Born theory (red curve).

8. RMS error table for the LINEARLY varying model.

9. Plots of all transfer free energy profiles for Born ions in all 9 co-solvent mixtures using a quadratically varying set of SLIC parameters as a function of concentration, compared to experiment (data points) and Born theory (red curve). To assess whether data sparsity affected model accuracy, we fit the transfer free energy profile for each ion to a quadratic, and allowed the quadratic fit to predict transfer free energies where experimental data were not available.

10. RMS error table for the quadratically varying model if we used additional data points in the parameterization (which were determined according to the procedure described above).

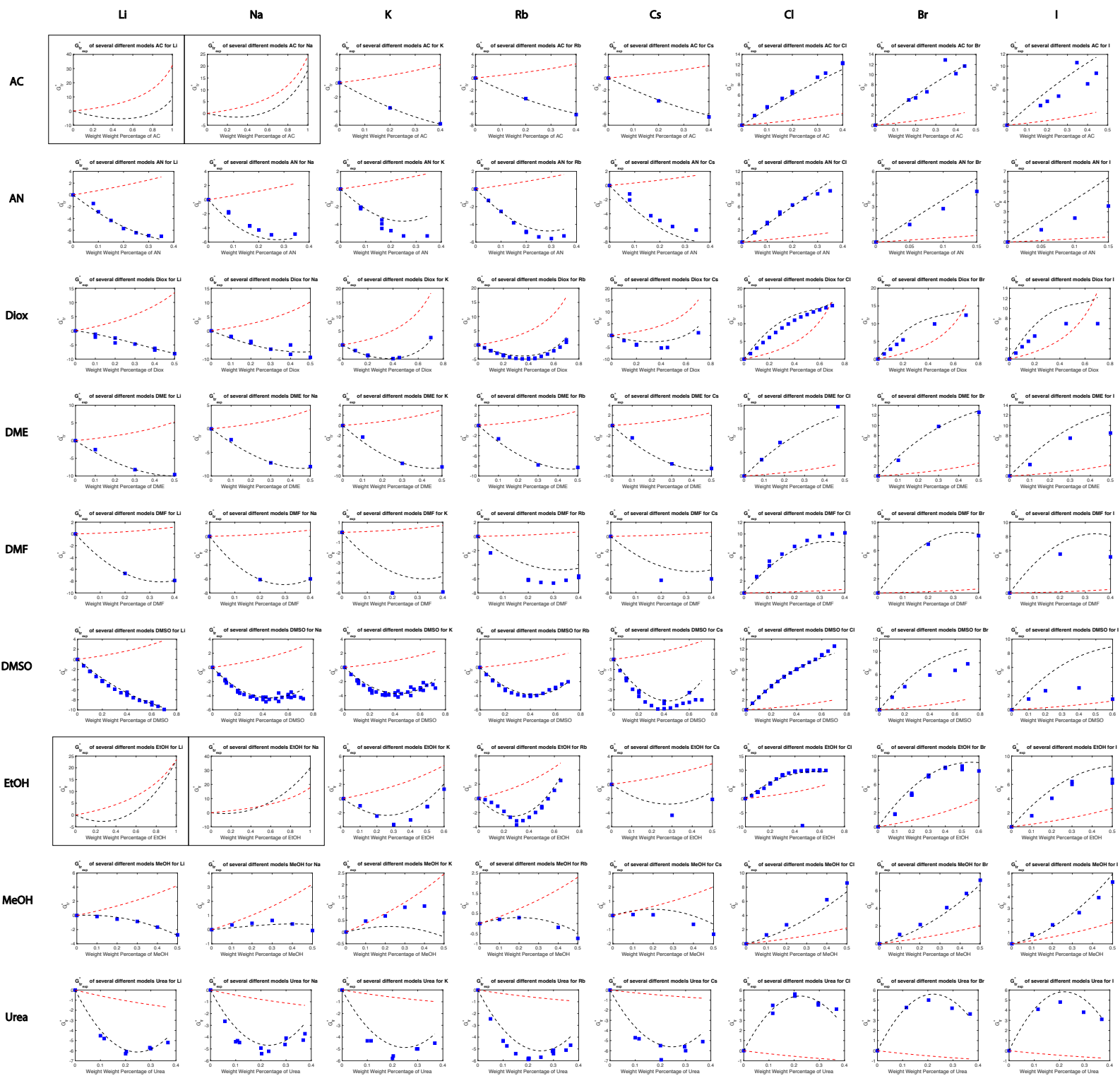
2

Figure showing ΔG° for Cations and Anions across solvents AC, AN, Diox, DME, DMF, DMSO, EtOH, MeOH, Urea as a function of concentration. Dashed lines represent SLIC model; markers represent experimental data for Li, Na, K, Rb, Cs (cations) and Cl, Br, I (anions). Values in kJ/mol.

|      | Li   | Na   | K    | Rb   | Cs   | Cl   | Br   | I    |
|------|------|------|------|------|------|------|------|------|
| **MeOH** | -529 | -423 | -351 | -333 | -303 | -305 | -277 | -242 |
| **DMSO** | -530 | -424 | -351 | -334 | -304 | -305 | -277 | -241 |
| **AC**   | -529 | -425 | -357 | -333 | -296 | -304 | -278 | -244 |
| **EtOH** | -527 | -421 | -354 | -331 | -294 | -304 | -278 | -245 |
| **AN**   | -529 | -423 | -351 | -333 | -302 | -305 | -278 | -243 |
| **Urea** | -529 | -424 | -351 | -333 | -303 | -305 | -278 | -241 |
| **DME**  | -529 | -424 | -356 | -333 | -296 | -304 | -278 | -244 |
| **Diox** | -530 | -423 | -350 | -332 | -302 | -307 | -279 | -242 |
| **DMF**  | -529 | -424 | -356 | -333 | -296 | -303 | -277 | -244 |

TABLE 1. Predicted $\Delta G_{solv}^{es}$ for Several Different Solvent-Ion Pairs in $\frac{kJ}{mol}$

| | MeOH - F | AC - Li | AC - Na | AC - F | DMSO - F | EtOH - Li | EtOH - Na | EtOH - F | AN - F | Urea - F | DME - F | Diox - F | DMF - F |
|---|---|---|---|---|---|---|---|---|---|---|---|---|---|
| 0 | 0 | 0 | 0 | 0 | 0 | 0 | 0 | 0 | 0 | 0 | 0 | 0 | 0 |
| 0.02 | 0.32 | -0.4 | -0.19 | 0.69 | 0.49 | -0.39 | -0.16 | 0.74 | 0.22 | 0.68 | 0.86 | 1.38 | 1.09 |
| 0.04 | 0.64 | -0.79 | -0.37 | 1.38 | 0.99 | -0.75 | -0.3 | 1.45 | 0.37 | 1.32 | 1.69 | 2.7 | 2.12 |
| 0.06 | 0.98 | -1.17 | -0.54 | 2.05 | 1.47 | -1.08 | -0.41 | 2.14 | 0.47 | 1.89 | 2.49 | 3.96 | 3.08 |
| 0.08 | 1.32 | -1.53 | -0.7 | 2.71 | 1.96 | -1.38 | -0.5 | 2.8 | 0.51 | 2.41 | 3.27 | 5.17 | 3.97 |
| 0.1 | 1.67 | -1.87 | -0.85 | 3.36 | 2.44 | -1.65 | -0.56 | 3.44 | 0.49 | 2.87 | 4.02 | 6.32 | 4.8 |
| 0.12 | 2.03 | -2.21 | -0.99 | 4.01 | 2.92 | -1.89 | -0.59 | 4.05 | 0.42 | 3.28 | 4.74 | 7.41 | 5.56 |
| 0.14 | 2.4 | -2.52 | -1.11 | 4.64 | 3.39 | -2.1 | -0.59 | 4.64 | 0.3 | 3.63 | 5.44 | 8.45 | 6.25 |
| 0.16 | 2.78 | -2.83 | -1.22 | 5.26 | 3.87 | -2.27 | -0.56 | 5.21 | 0.12 | 3.92 | 6.12 | 9.44 | 6.87 |
| 0.18 | 3.17 | -3.11 | -1.32 | 5.87 | 4.33 | -2.42 | -0.51 | 5.74 | -0.11 | 4.15 | 6.76 | 10.38 | 7.43 |
| 0.2 | 3.57 | -3.39 | -1.4 | 6.48 | 4.8 | -2.54 | -0.43 | 6.26 | -0.4 | 4.33 | 7.38 | 11.27 | 7.93 |
| 0.22 | 3.97 | -3.64 | -1.47 | 7.07 | 5.26 | -2.62 | -0.32 | 6.75 | -0.74 | 4.45 | 7.98 | 12.11 | 8.35 |
| 0.24 | 4.39 | -3.89 | -1.52 | 7.65 | 5.72 | -2.68 | -0.18 | 7.21 | -1.13 | 4.52 | 8.55 | 12.9 | 8.72 |
| 0.26 | 4.82 | -4.11 | -1.56 | 8.23 | 6.17 | -2.7 | -0.01 | 7.66 | -1.59 | 4.53 | 9.1 | 13.64 | 9.01 |
| 0.28 | 5.26 | -4.32 | -1.58 | 8.8 | 6.62 | -2.69 | 0.19 | 8.07 | -2.11 | 4.48 | 9.63 | 14.34 | 9.24 |
| 0.3 | 5.7 | -4.51 | -1.58 | 9.36 | 7.07 | -2.65 | 0.42 | 8.47 | -2.7 | 4.38 | 10.13 | 15 | 9.4 |
| 0.32 | 6.16 | -4.69 | -1.57 | 9.91 | 7.52 | -2.57 | 0.68 | 8.84 | -3.36 | 4.23 | 10.61 | 15.61 | 9.5 |
| 0.34 | 6.63 | -4.84 | -1.54 | 10.46 | 7.96 | -2.47 | 0.97 | 9.19 | -4.11 | 4.03 | 11.06 | 16.19 | 9.53 |
| 0.36 | 7.11 | -4.98 | -1.49 | 10.99 | 8.41 | -2.33 | 1.3 | 9.51 | -4.95 | 3.78 | 11.49 | 16.73 | 9.5 |
| 0.38 | 7.6 | -5.1 | -1.43 | 11.52 | 8.85 | -2.16 | 1.65 | 9.82 | -5.88 | 3.48 | 11.91 | 17.23 | 9.4 |
| 0.4 | 8.1 | -5.2 | -1.34 | 12.05 | 9.29 | -1.95 | 2.04 | 10.1 | -6.93 | 3.13 | 12.3 | 17.7 | 9.24 |
| 0.42 | 8.61 | -5.29 | -1.23 | 12.57 | 9.72 | -1.71 | 2.46 | 10.35 | -8.1 | 2.73 | 12.67 | 18.14 | 9.01 |
| 0.44 | 9.14 | -5.35 | -1.1 | 13.08 | 10.16 | -1.44 | 2.91 | 10.59 | -9.4 | 2.28 | 13.02 | 18.56 | 8.71 |
| 0.46 | 9.68 | -5.39 | -0.95 | 13.59 | 10.6 | -1.13 | 3.39 | 10.81 | -10.84 | 1.78 | 13.35 | 18.95 | 8.35 |
| 0.48 | 10.22 | -5.41 | -0.77 | 14.1 | 11.04 | -0.78 | 3.91 | 11 | -12.44 | 1.24 | 13.67 | 19.32 | 7.93 |
| 0.5 | 10.78 | -5.4 | -0.58 | 14.6 | 11.47 | -0.4 | 4.46 | 11.18 | -14.2 | 0.65 | 13.97 | 19.69 | 7.44 |
| 0.52 | 11.36 | -5.37 | -0.35 | 15.1 | 11.91 | 0.01 | 5.05 | 11.33 | -16.15 | 0.01 | 14.25 | 20.04 | 6.89 |
| 0.54 | 11.94 | -5.32 | -0.1 | 15.6 | 12.35 | 0.46 | 5.67 | 11.46 | -18.29 | -0.67 | 14.53 | 20.39 | 6.28 |
| 0.56 | 12.54 | -5.24 | 0.18 | 16.1 | 12.79 | 0.95 | 6.33 | 11.58 | -20.64 | -1.41 | 14.79 | 20.75 | 5.6 |
| 0.58 | 13.15 | -5.14 | 0.48 | 16.6 | 13.22 | 1.48 | 7.03 | 11.68 | -23.2 | -2.19 | 15.04 | 21.13 | 4.86 |
| 0.6 | 13.78 | -5.01 | 0.82 | 17.1 | 13.66 | 2.05 | 7.76 | 11.75 | -26 | -3.03 | 15.28 | 21.54 | 4.05 |

TABLE 2. Predicted $\Delta G_{tr}^\circ$ for Unknown Ions in Different Mixtures in $\frac{kJ}{mol}$

|     | K    | Rb   | Cs   | Cl   | Br   | I    |
| --- | ---- | ---- | ---- | ---- | ---- | ---- |
| **0**   | 1.94 | 2.07 | 2.33 | 2.25 | 2.46 | 2.8  |
| **0.1** | 1.93 | 2.06 | 2.32 | 2.28 | 2.49 | 2.83 |
| **0.2** | 1.92 | 2.05 | 2.31 | 2.29 | 2.51 | 2.86 |
| **0.3** | 1.92 | 2.05 | 2.3  | 2.3  | 2.52 | 2.87 |
| **0.4** | 1.92 | 2.05 | 2.3  | 2.31 | 2.53 | 2.88 |
| **0.5** | 1.92 | 2.05 | 2.31 | 2.31 | 2.53 | 2.88 |
| **0.6** | 1.93 | 2.06 | 2.32 | 2.3  | 2.52 | 2.87 |
| **0.7** | 1.94 | 2.07 | 2.33 | 2.29 | 2.51 | 2.86 |
| **0.8** | 1.95 | 2.09 | 2.35 | 2.28 | 2.49 | 2.83 |
| **0.9** | 1.97 | 2.11 | 2.38 | 2.26 | 2.46 | 2.8  |
| **1**   | 2    | 2.14 | 2.41 | 2.23 | 2.43 | 2.76 |

TABLE 3. Predicted Born Radii for EtOH-W Mixtures

|         | $\Delta G_{solv}$ |        | $\Delta G^{tr}$ |        |
|---------|---------|--------|---------|--------|
|         | **Cations** | **Anions** | **Cations** | **Anions** |
| **MeOH** | 2.24 | 1.5  | 0.41 | 0.54 |
| **DMSO** | 2.22 | 2.58 | 0.47 | 1.71 |
| **AC**   | 6.64 | 1.1  | 0.13 | 1.38 |
| **EtOH** | 4.45 | 2.41 | 0.86 | 2.46 |
| **AN**   | 2.25 | 1.27 | 0.86 | 0.95 |
| **Urea** | 2.48 | 1.63 | 0.72 | 0.53 |
| **DME**  | 5.07 | 1.14 | 0.42 | 1.5  |
| **Diox** | 2.71 | 1.95 | 1.09 | 1.81 |
| **DMF**  | 4.62 | 0.75 | 1.29 | 1.07 |

TABLE 4. RMS Error for $\Delta G_{solv}^{es}$ and $\Delta G_{tr}^{\circ}$ Using Quadratically Varying Model Parameters for Cations and Anions in Several Different Mixtures

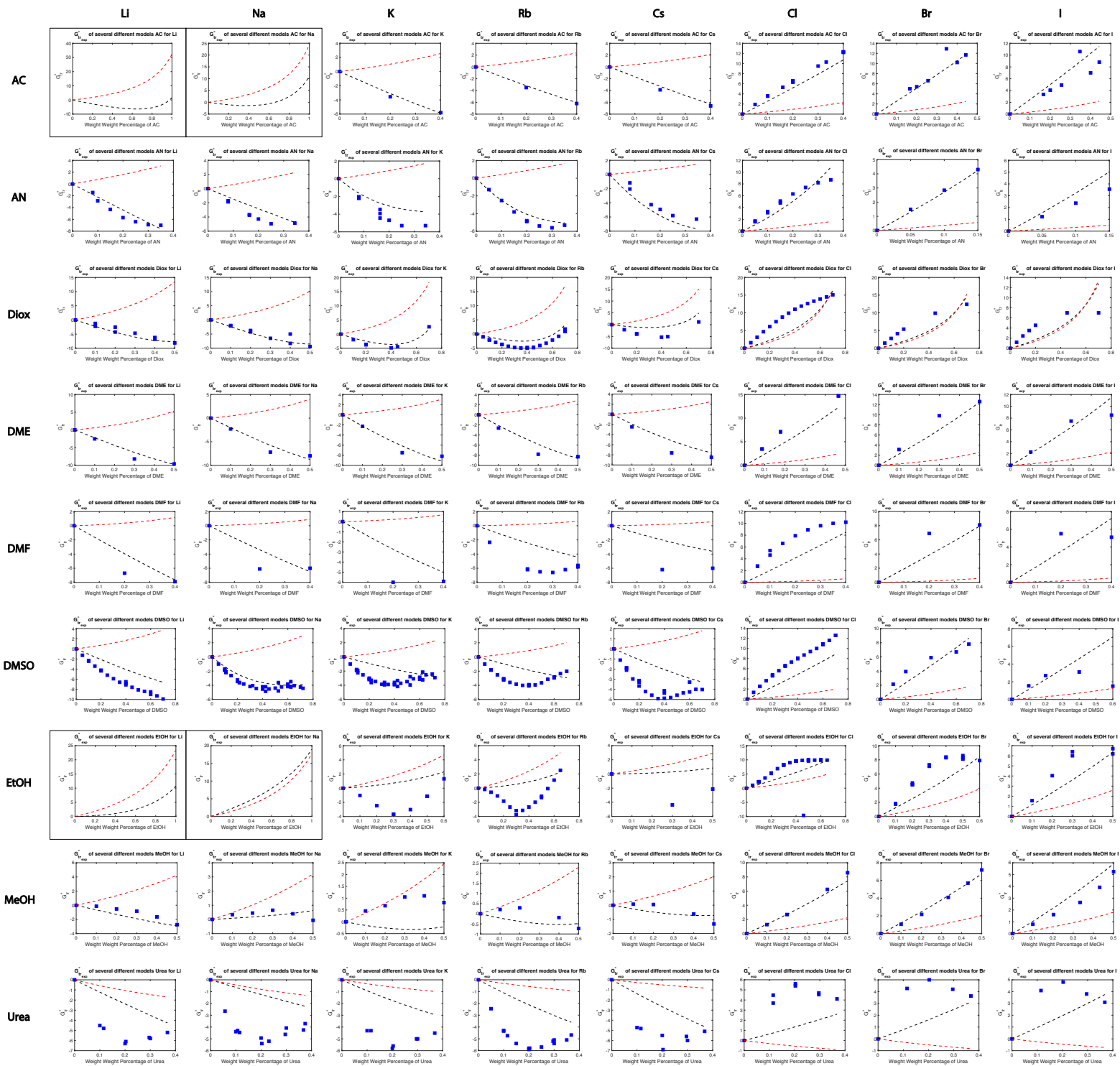

|  | $\Delta G_{solv}$ | | $\Delta G^{tr}$ | |
|---|---|---|---|---|
|  | **Cations** | **Anions** | **Cations** | **Anions** |
| **MeOH** | 2.25 | 1.54 | 0.63 | 0.55 |
| **DMSO** | 4.93 | 1.8 | 1.71 | 2.54 |
| **AC** | 6.64 | 1.14 | 0.3 | 1.51 |
| **EtOH** | 4.59 | 2.61 | 2.77 | 3.23 |
| **AN** | 2.3 | 1.32 | 0.93 | 0.91 |
| **Urea** | 2.72 | 2.21 | 2.86 | 2.54 |
| **DME** | 2.62 | 2.19 | 1 | 1.83 |
| **Diox** | 2.81 | 3.02 | 1.8 | 3.29 |
| **DMF** | 2.92 | 2.15 | 2.29 | 2.46 |

TABLE 5. RMS Error for $\Delta G_{solv}^{es}$ and $\Delta G_{tr}^{\circ}$ Using Linearly Varying Model Parameters for Cations and Anions in Several Different Mixtures

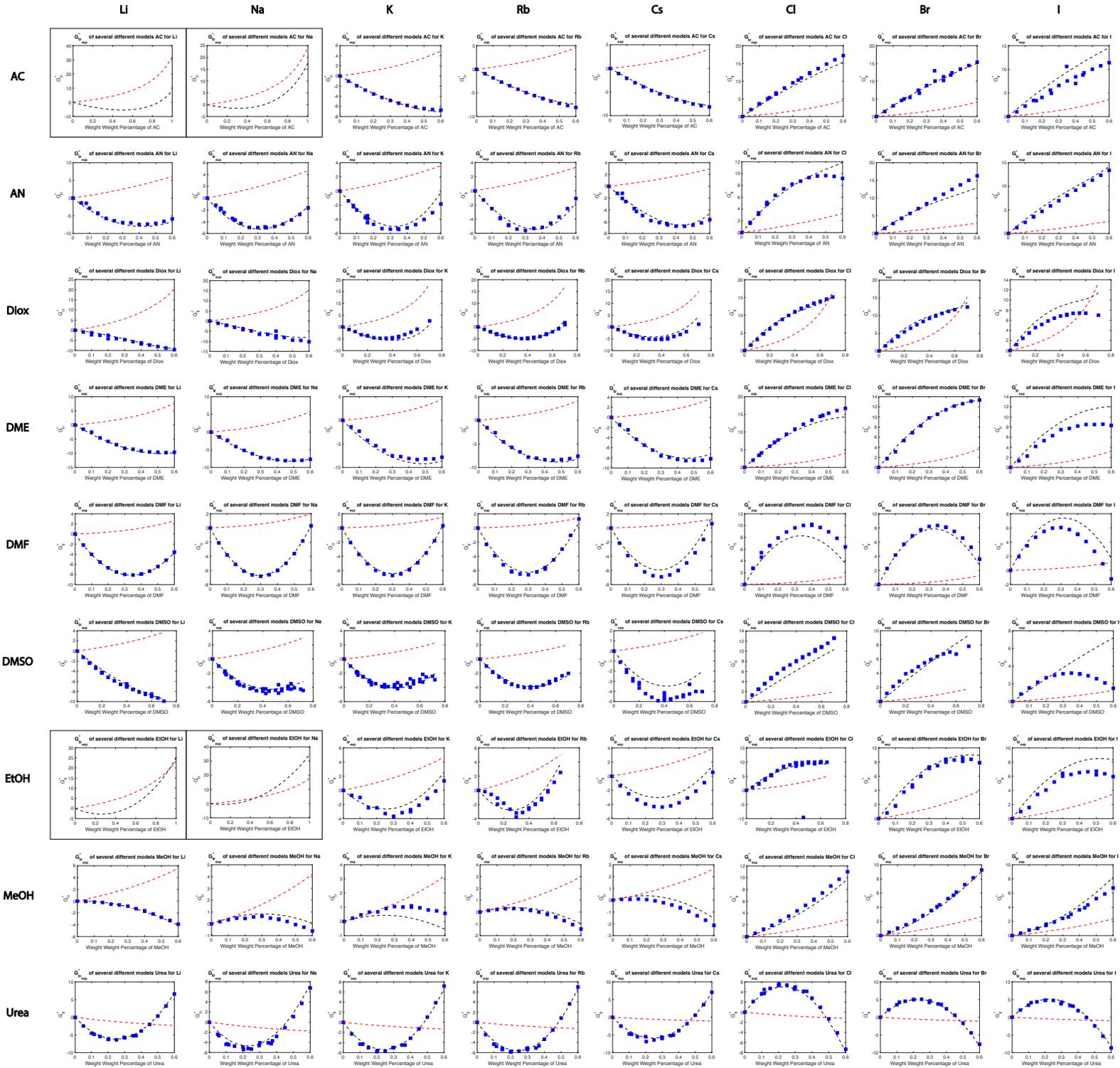

- - - SLIC   ☐ SLIC without Experimental Data   - - - Born   ■ Experiment   in kJ/mol

10

|        | $\Delta G_{solv}$ | | $\Delta G^{tr}$ | |
|--------|---------|--------|---------|--------|
|        | Cations | Anions | Cations | Anions |
| **MeOH** | 2.36 | 1.63 | 0.41 | 0.61 |
| **DMSO** | 2.35 | 2.69 | 0.56 | 1.73 |
| **AC**   | 6.66 | 1    | 0.22 | 1.41 |
| **EtOH** | 6.43 | 2.13 | 0.94 | 2.22 |
| **AN**   | 2.39 | 1.55 | 0.51 | 1.09 |
| **Urea** | 2.45 | 1.41 | 0.5  | 0.37 |
| **DME**  | 2.5  | 2.26 | 0.45 | 1.38 |
| **Diox** | 2.76 | 2.16 | 0.93 | 1.21 |
| **DMF**  | 2.63 | 2.23 | 0.43 | 1.4  |

TABLE 6. RMS Error for $\Delta G_{solv}^{es}$ and $\Delta G_{tr}^{\circ}$ Using Quadratically Varying Model Parameters and Interpolated Data Points for Cations and Anions in Several Different Mixtures



\end{document}